\documentclass[aps,pra,superscriptaddress,twocolumn,floatfix,a4paper]{revtex4}

\usepackage{graphicx,graphics,epsfig}   
\usepackage{dcolumn}    
\usepackage{bm}         
\usepackage{amsmath}    
\usepackage{verbatim}   
\usepackage{color}      
\usepackage{subfigure}  
\usepackage{times,natbib}
\usepackage{amsmath,amsfonts,amssymb,graphics,graphics,color,times}

\usepackage{latexsym}
\usepackage{amsmath}
\usepackage{amssymb}
\usepackage{amsfonts}
\usepackage{amsthm}
\usepackage{mathrsfs}
\usepackage{color,verbatim,graphics}
\usepackage{psfrag}
\DeclareMathAlphabet{\mathrsfs}{U}{rsfs}{m}{n}
\DeclareMathAlphabet{\mathpzc}{OT1}{pzc}{m}{it}
\DeclareMathAlphabet{\matheus}{U}{eus}{m}{n}
\DeclareMathAlphabet{\mathbbold}{U}{bbold}{m}{n}

\newcommand{\ba}{\begin{eqnarray}}
\newcommand{\ea}{\end{eqnarray}}
\newcommand{\ban}{\begin{eqnarray*}}
\newcommand{\ean}{\end{eqnarray*}}

\newcommand{\ket}[1]{|#1\rangle}
\newcommand{\bra}[1]{\langle#1|}

\definecolor{nblue}{rgb}{0.2,0.2,0.7}

\begin{document}

\title{Persistency of entanglement and nonlocality in multipartite quantum systems}

\author{Nicolas Brunner}
\affiliation{H.H. Wills Physics Laboratory, University of Bristol, Tyndall Avenue, Bristol, BS8 1TL, United Kingdom}
\affiliation{D\'epartement de Physique Th\'eorique, Universit\'e de Gen\`eve, 1211 Gen\`eve, Switzerland}
\author{Tam\'as V\'ertesi}
\affiliation{Institute of Nuclear Research of the Hungarian Academy of Sciences,
H-4001 Debrecen, P.O. Box 51, Hungary}



\begin{abstract}
The behaviour under particle loss of entanglement and nonlocality is investigated in multipartite quantum systems. In particular, we define a notion of persistency of nonlocality, which leads to device-independent tests of persistent entanglement. We investigate the persistency of various classes of multipartite quantum states, exhibiting a variety of different behaviours. A particular attention is devoted to states featuring maximal persistency.
Finally we discuss a link between the symmetry of a state and its persistency, illustrating the fact that too much symmetry reduces the strength of correlations among subsystems. These ideas also lead to a device-independent estimation of the asymmetry of a quantum state.
\end{abstract}

\maketitle

Entanglement and nonlocality are now recognized as central aspects of quantum mechanics, and play a prominent role in quantum information. Separated entangled quantum systems are correlated in a genuine quantum manner, admitting no analogue in classical physics \cite{RMP_horo,guhne}.
The strongest manifestation of such quantum correlations is nonlocality, witnessed by the violation of a Bell inequality \cite{bell}.

In recent years, an intense research effort has been devoted to understanding quantum correlations. While significant progress was made for a scenario involving two systems, it is fair to say that the multipartite case remains poorly understood. This is nevertheless highly desirable given the importance of multipartite quantum correlations in many-body physics \cite{osterloh} and in quantum information, for quantum networks \cite{acin,kimble}, quantum computation \cite{MBQC} and quantum memories \cite{Qmem}. Moreover, the creation of complex multipartite entangled state is now achievable experimentally in various types of systems, including photons \cite{pan8}, ions \cite{blatt}, and superconducting qubits \cite{neeley}.

A main obstacle to a deep understanding of multipartite quantum correlations is indeed the complex structure of entanglement when three or more parties are involved \cite{dur}. However, another fundamental issue is arguably the lack of adapted tools for characterizing multi-particle correlations. So far, most efforts were devoted to investigate the notion of genuine multipartite entanglement (GME) \cite{RMP_horo,guhne}. Loosely speaking, a state is said to be genuine multipartite entangled if all its parties are genuinely entangled together, that is, no subgroups can be factorized out. Whereas GME definitely captures an important aspect of multipartite entanglement, it fails to capture many other relevant features. In particular, GME is unrelated to robustness to particle losses. Consider for instance GHZ and W states, both GME; in the former case, all entanglement vanishes upon loss of a single particle while in the second case, almost all particles must be lost before entanglement disappears. Notably, robustness to losses has important implications from a physical point of view. Consider for instance Dicke states \cite{dicke} (including W states), central to the physics of the interaction of light and matter, that are highly robust to losses, hence ideal for applications such as quantum memories \cite{Qmem}. More generally, quantum states with robust entanglement and nonlocality are attractive for a wide range of applications in quantum information. Thus, better understanding the behaviour under losses of multipartite quantum correlations, and developing tools to characterize this process, is of both fundamental and applied significance.

Here we investigate the robustness of entanglement and nonlocality under particle loss. In particular, we define in Section~I a notion of persistency of nonlocality---loosely speaking, the minimal number of particles to be lost for nonlocal correlations to vanish. This provides a way for testing persistent entanglement in a device-independent manner \cite{DI}, that is without placing any assumption about the measurement devices and on the Hilbert space dimension of the quantum state. In Sections~\ref{cs} and IV we investigate the persistency of quantum states with specific symmetries, including cluster states \cite{briegel}, the resource for measurement based quantum computation \cite{MBQC}. In Section~III we discuss states having maximal persistency. Finally, after commenting on the relation between persistency and GME in Section~V, we discuss in Section~VI the interplay between the symmetry of a state and its persistency, which indicates that too much symmetry weakens nonlocal correlations among subsystems.
These ideas lead to a device-independent test of the asymmetry of quantum states. We conclude in Section~VII.


\section{Persistency}

Consider a quantum state $\rho$ of $N$ systems. Take the partial trace over $k<N$ systems $j_1,...,j_k\in \{1,...,N\}$, and denote the reduced state \ba \rho_{(j_1,...,j_k)} = \text{tr}_{j_1,...,j_k} (\rho) \ea
The strong persistency of entanglement of $\rho$, $P_E(\rho)$, is then defined as
the minimal $k$ such that the reduced state $ \rho_{(j_1,...,j_k)}$ becomes fully separable, for at least one set of subsystems $j_1,...,j_k $.

The persistency of nonlocality of $\rho$, $P_{NL}(\rho)$, is defined in a similar way, but now demanding only that the reduced state $ \rho_{(j_1,...,j_k)}$ becomes local, i.e. that the probability distribution obtained from local measurements on $ \rho_{(j_1,...,j_k)}$ does not violate any Bell inequality. Formally this means that the probability distribution
\ba p(a_1...a_{N}|x_1...x_{N}) = \text{tr}(\rho M_{a_1}^{x_1}\otimes ... \otimes M_{a_N}^{x_N})\ea
admits a hidden variable model for general local measurement operators $M_{a_i}^{x_i}$, with $M_{a_i}^{x_i}=\openone$ if $i=j_1,...,j_k$ (the systems that have been traced out) and $\sum_{a_i} M_{a_i}^{x_i}=\openone$ otherwise. Here $x_i$ and $a_i$ denote the measurement setting and its outcome, respectively, of party $i$.
Note that we will also be lead to consider the concept of hidden nonlocality \cite{sandu} in this context. That is, we strengthen the above definition and demand that the reduced state $ \rho_{(j_1,...,j_k)}$ is local even after the remaining parties have performed a local filtering. In this case, persistency of nonlocality is denoted by $P^*_{NL}(\rho)$.

It is instructive to compare these notions of persistency, and also to relate them with the slightly different concept of persistency of entanglement ($p_E$) introduced in Ref.~\cite{briegel}; $p_E$ is the minimal number of parties $k$ that must make local measurements on their subsystems, such that the state of the remaining $N-k$ parties is separable, for all possible measurement outcomes. For any state $\rho$, one has that
\ba N-1\geq p_{E}(\rho) \geq P_E(\rho) \geq P^*_{NL}(\rho) \geq P_{NL}(\rho) \geq 1. \ea
The second inequality follows from the fact that (i) taking the partial trace over a given system is equivalent to performing a local measurement on this system and then forgetting the measurement outcome, and (ii) the mixture of separable states is obviously separable, whereas the mixture of entangled states is not necessarily entangled. The third inequality comes from the fact that (i) entanglement is necessary for having quantum nonlocality, and (ii) there exist entangled states which are local \cite{werner}. The fourth inequality follows from the fact that there exist local quantum states featuring hidden nonlocality \cite{sandu}.

The persistencies of entanglement ($P_E$) and nonlocality ($P_{NL}$) \footnote{From now on $P_E$ will be simply referred to as persistency of entanglement.} represent the minimum number of parties that must be traced out from $\rho$ so that the reduced state becomes separable, and local, respectively. These notions thus have a clear operational meaning, characterizing the robustness to particle loss of multipartite quantum correlations. In particular, the persistency of entanglement has been investigated experimentally for simple multi-qubit states \cite{mohamed}. In the following we will focus mostly on $P_{NL}(\rho)$ which is of both conceptual and experimental interest, since it represents a lower bound on $P_E(\rho)$ (hence also on $p_E(\rho)$) that can be derived in a device-independent way.
Clearly, if all reduced states of $\rho$ violate a Bell inequality when $k$ parties are traced out, then all reduced states must be entangled, irrespectively of the Hilbert space dimension of the state and the alignment of the measurement devices.

Below we investigate $P_E$ and $P_{NL}$ for some of the main classes of multipartite quantum states, revealing a range of different behaviours. As mentioned above, there exist GME states which have minimal persistency (i.e. $P_E=1$); this is the case for GHZ states, as well for their generalizations to an arbitrary number of parties and dimensions. More interesting behaviours will be described in the next sections. First, we will see that for cluster states, the persistencies grow linearly with the number of subsystems $N$: $P_{E},P_{NL}\sim \alpha N$ with $\alpha \in [1/4, 2/5]$. Then, we will see that there are states featuring maximal persistency of both entanglement and nonlocality, i.e. $P_E=P_{NL}=N-1$.

 \begin{figure}[b!]
  \includegraphics[width=0.75\columnwidth]{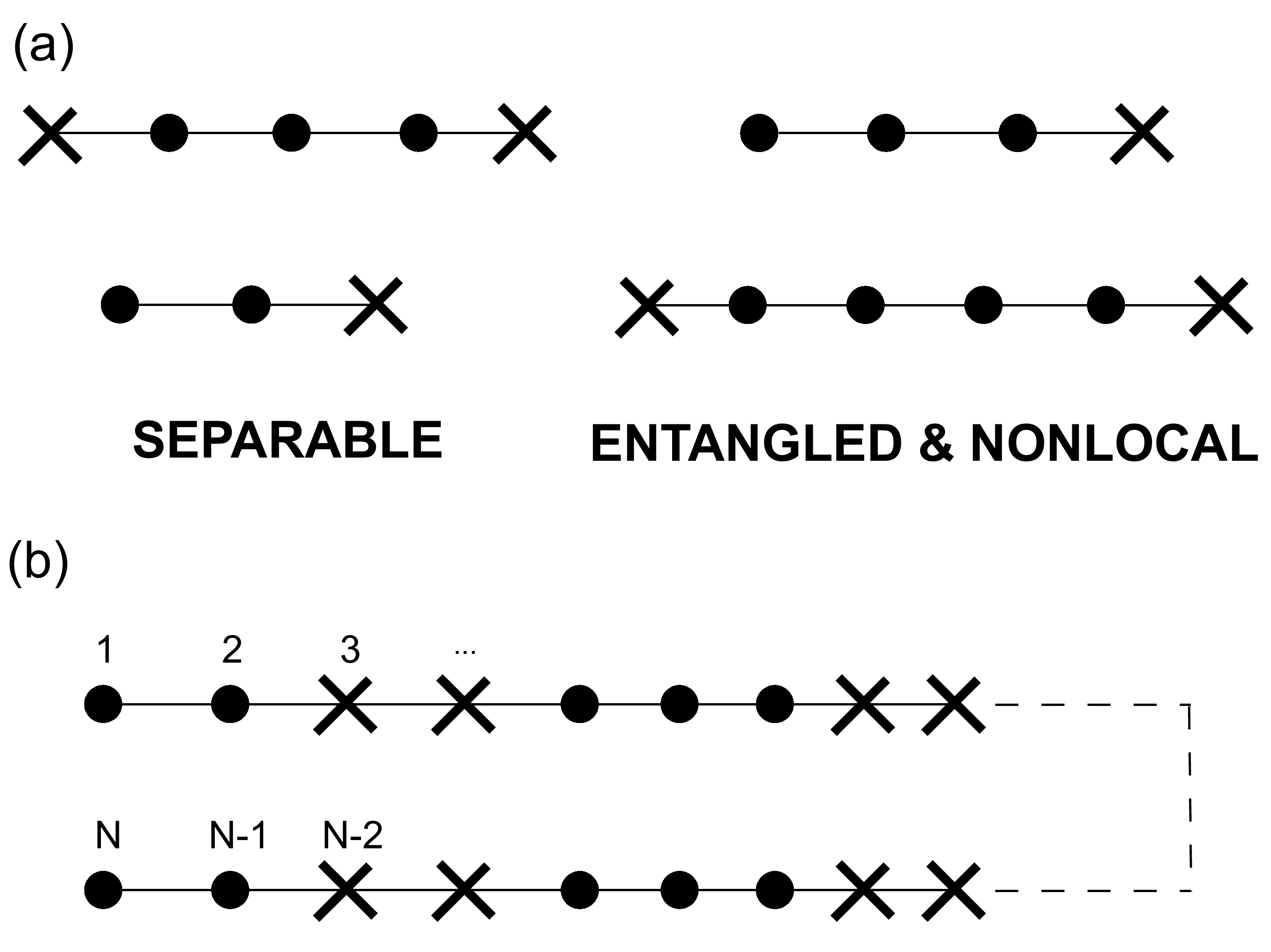}
  \caption{Persistency of 1D cluster states. (a) Small portions of a 1D chain which are separable or entangled and nonlocal. (b) Procedure to fully suppress correlations in a chain. Note that two consecutive qubits must be traced out in order to guarantee all correlations (including classical ones) are removed.}
\label{fig}
\end{figure}

\section{Cluster states} \label{cs}

We first consider 1D cluster states, in particular closed chains of qubits with Ising type interactions between next neighbours:
\ba \ket{L_N} = \frac{1}{2^{N/2}}  \bigotimes_{a=1}^N (\ket{0}_a \sigma_z^{a+1}+\ket{1}_a) \ea
where $\sigma_z^{N+1}=1$.

For $N=2,3$ the state is equivalent (up to local unitaries) to a Bell state and to a (3-qubit) GHZ state, respectively. Hence we have that $P_E=P_{NL}=1$.

Hence, let us start with $\ket{L_4}$.
Tracing out any of the 4 qubits gives a reduced state of the form
\ba \rho = \ket{0}\bra{0} \otimes \ket{L_2}\bra{L_2} + \ket{1}\bra{1} \otimes \ket{L'_2}\bra{L'_2} \ea
where $\ket{L'_2}=\sigma_z^1 \ket{L_2}$. State $\rho$ is clearly entangled, as a local measurement on the first qubit will project the remaining two on a maximally entangled state. Tracing out a second qubit results in a separable state. Note that $\rho$ is also nonlocal, and violates a simple tripartite Bell inequality based on the Clauser-Horne-Shimony-Holt expression \cite{chsh}: 
\ba\label{BI} I=(1+A) \text{CHSH}_{BC} + 2(1-A) \leq 4 \ea
where $\text{CHSH}_{BC}=B_0C_0+B_0C_1+B_1C_0-B_1C_1$ denotes the Clauser-Horne-Shimony-Holt Bell expression \cite{chsh}; here $B_j$ denotes the $j$-th measurement of party $B$, with binary outcomes $\pm1$; each term, e.g. $B_jC_k$ should be understood as the expectation value of the product of the outcomes. Suitable local measurements on $\rho$ give $I=4\sqrt{2}$. Note the unusual feature of the Bell inequality \eqref{BI}, namely that party A performs always the same measurement which acts as a filtering, heralding the desired entangled state for the remaining two parties. Since the parties are not allowed to communicate, \eqref{BI} is indeed a valid Bell inequality.

For a chain of an arbitrary length, we obtain bounds on the persistencies.
In order to get upper bounds on $P_E$, we determine basically in how many parts the chain must be broken in order to remove all entanglement. Lower bounds on $P_{NL}$ are obtained by determining the minimum chain section that is nonlocal.
The relevant results are summarized in Fig.~1a.
On the one hand, considering a chain of 6 qubits, and tracing qubits 1 and 6 results in a state of the form
\ba \rho &=& \ket{L_4}\bra{L_4} + \sigma_z^1\ket{L_4}\bra{L_4}\sigma_z^1 \\ \nonumber &  +& \sigma_z^4\ket{L_4}\bra{L_4}\sigma_z^4
+ \sigma_z^1 \sigma_z^4\ket{L_4}\bra{L_4}\sigma_z^1 \sigma_z^4 .\ea
It can be checked that by projecting qubits 1 and 4 of $\rho$ prepares qubits 2 and 3 in a qubit Bell pair. Hence the above state $\rho$ is entangled. Moreover, it is nonlocal, violating an inequality similar to \eqref{BI}. Thus, for a chain of arbitrary length, whenever 4 (or more) neighbouring qubits are left untouched, the reduced state is entangled and nonlocal.
On the other hand, considering a chain of 5 qubits and tracing out qubits 1 and 5 results in a separable state of the form $ \rho = \frac{1}{8}(\openone^{\otimes 3}+ \sigma_z^{\otimes 3})$.
Hence, in a chain, if all subsets of neighbouring parties are no larger than three qubits, then the reduced state is separable.
Note however that here, in order to fully disentangle a chain, we always trace out two neighbouring qubits in order to remove all correlations (including classical ones) between the remaining parts of the chain (see Fig.~1b). We believe that this method is too conservative in general, and that less qubits could be traced out. For the ring cluster $\ket{R_N}$ (identical to $\ket{L_N}$, but with $\sigma_z^{N+1}=\sigma_z$) we obtain
\ba\label{bounds} \lfloor \frac{N-1}{4} \rfloor +1 \leq P_{NL}\leq P_{E} \leq 2 \lceil \frac{N}{5} \rceil. \ea
For the linear cluster, we get slightly different bounds:
\ba \lfloor \frac{N-1}{4} \rfloor +1 \leq P_{NL} \leq P_{E} \leq 2 \lceil \frac{N-6}{5} \rceil+2 \ea
due to the edges of the chain.
Therefore, in the asymptotic regime the persistencies grow linearly with $N$: $P_{E},P_{NL}\sim \alpha N$ with $\alpha \in [1/4, 2/5]$---note that $p_E\sim N/2$ \cite{briegel}.

Finally, it is worth noting that both persistencies have roughly the same behaviour. Hence, after tracing out almost $P_E$ parties, the remaining entanglement is still pretty strong, as it leads to nonlocality.


    \begin{table*}
    \caption{Persistencies of entanglement $P_E$ and nonlocality $P_{NL}$ and the strength $w$ of $P_{NL}$ are shown for various classes of states up to $N=7$ qubits, including W states, Dicke states, translationally invariant states, ring and linear cluster states. For $N=6$, we considered 2D cluster states: $\ket{Cl_{2\times3}}$ and $\ket{Cl^p_{2\times3}}$ (see \cite{jung}). All values of $P_E$ are optimal; values of $P_{NL}$ are optimal only when $P_E=P_{NL}$.
    }\label{table}
    \centering
    \begin{tabular}{ c | c c c c c c c c c | c c c c c c}
    \hline \hline
    State &  $\ket{W_6}$ &  $\ket{D_6^2}$ & $\ket{D_6^3}$ & $\ket{T_6^2}$ & $\ket{T_6^3}$ & $\ket{L_6}$ & $\ket{R_6}$ & $\ket{Cl^p_{2\times3}}$ & $\ket{Cl_{2\times3}}$ &  $\ket{W_7}$ & $\ket{D_7^3}$ & $\ket{T_7^3}$ & $\ket{R_7}$ & $\ket{L_7}$  \\
    \hline  \\
    $P_E$ & 5 & 5 & 5 & 4 & 3 & 2 & 3 & 3 & 3 & 6 & 6 & 4 & 3 & 3 & \\
    $P_{NL}$ & 2 & 2 & 3 & 3 & 3 & 2 & 3 & 3 & 3 & 3 & 3 & 3 & 3 & 3 & \\
    $w$ & 0.751 & 0.783 & 0.978 & 0.644 & 0.644 & 0.547 & 0.707 & 0.667 & 0.707 & 0.985 & 0.968 & 0.514 & 0.667 & 0.707 & \\
    \hline \hline
    \end{tabular}

    \vspace{3 mm}

    \centering
    \begin{tabular}{ c | c | c c c c | c c c c c}
    \hline \hline
    State &  $\ket{W_3}$ &  $\ket{W_4}$ & $\ket{D_4^2}$ & $\ket{T_4^2}$ & $\ket{L_4}$ & $\ket{W_5}$ & $\ket{D_5^2}$ & $\ket{T_5^2}$ & $\ket{L_5}$ &  $\ket{R_5}$ \\
    \hline  \\
    $P_E$ & 2 & 3 & 3 & 2 & 2 & 4 & 4 & 3 & 2 & 2  \\
    $P_{NL}$ & 1 & 2 & 1 & 2 & 2 & 2 & 2 & 3 & 2 & 2  \\
    $w$ & 0.644 & 0.989 & 0.471 & 0.707 & 0.707 & 0.860 & 0.907 & 0.772 & 0.667 & 0.577 \\
    \hline \hline
    \end{tabular}
    \end{table*}


\section{Maximal persistency}

Of interest are states featuring maximal persistency, hence being extremely robust against losses. We start by considering W states of $N$ particles:
\ba\label{W}    \ket{W_N} = \frac{1}{\sqrt{N}} \left( \ket{0 \hdots 0 1} +   \ket{0 \hdots 1 0} + \hdots + \ket{1 0 \hdots  0} \right)  \ea
which are proven to have maximal persistency of entanglement $P_E=N-1$ \cite{dur}. We will now see that the W state also has maximal persistency of nonlocality when hidden nonlocality is allowed, i.e. $P^*_{NL}=N-1$.
Consider the reduced state of any two parties of $\ket{W_N}$, which is of the form
\ba \rho(p) = p \ket{W_2}\bra{W_2} + (1-p) \ket{00}\bra{00} \ea
with $p=2/N$. Both parties now perform a local filtering, which can help to reveal the hidden nonlocality of certain quantum states \cite{sandu,W3}. Here we choose a filtering of the form
\ba T=\left(\begin{array}{ccc}
\epsilon & 0\\
0 & 1\\
\end{array}\right), \ea
where $0<\epsilon\le 1$ represents the attenuation of the state $\ket{0}$. After filtering, $\rho(p)$ becomes
\ba \rho_{f} = \frac{p}{(1-p)\epsilon^2+p}\ket{W_2}\bra{W_2} +
\frac{(1-p)\epsilon^2}{(1-p)\epsilon^2+p}\ket{00}\bra{00}. \nonumber\ea
Using the Horodecki criterion \cite{horoCHSH}, one gets
the maximum CHSH value of $\rho_f$ as a function of $p$ and $\epsilon$: \ba
\text{CHSH}(\rho_f) = \frac{2\sqrt 2p }{(1-p)\epsilon^2+p}.\ea
By choosing $\epsilon$ small enough this value can be brought arbitrary close to Tsirelson's bound of
$\text{CHSH}=2\sqrt{2}$ \cite{tsirelson}, for any $p$. 
Hence the W state has maximal persistency of nonlocality $P^*_{NL}=N-1$ for all $N$.
It would be interesting to determine the behaviour of $P_{NL}$ for the W state, that is, in the case where hidden nonlocality is not considered. We conjecture that  $P_{NL}<N-1$, and that the state $\rho(p)$ admits a local model for some values of $p>0$.
To support this conjecture, we note that $\rho(2/N)$ cannot violate any Bell inequality with less than $N$ settings per party, as the state admits an $(N-1)$-symmetric extension \cite{terhal}; note that this symmetric extension is given by the W state itself.

More generally, this raises the question of finding quantum states featuring maximal persistency $P_{NL}=N-1$.
Such states have the appealing feature that any two parties share nonlocal correlations with unit probability even if all other subsystems have been lost.
A trivial example of such a state is a 'fully connected Bell state', that is, an N-party state for which any two parties share a 2-qubit Bell pair. Note however that such states are extremely costly in terms of local Hilbert space dimension: $d=2^{N-1}$.

It is then natural to ask whether there exist lower-dimensional states with the same property.
We construct now an $N$-partite state $\ket{\Psi_N}$ ($N$ is odd) with local dimension $d=N$ with $P_{NL}(\Psi_N)=N-1$. 
Below we will present the state for the simplest case $N=3$; the general construction can be found in the Appendix B. Consider the translationally invariant state
\ba\label{3qutrit} \ket{\Psi_3} = a \ket{000} + b (\ket{012}+\ket{201}+\ket{120}) \ea
After tracing out any of the subsystems, we get (up to relabeling) the 2-party reduced state
\ba \rho = (1-2b^2) \ket{\psi_\theta}\bra{\psi_\theta} + b^2 (\ket{02}\bra{02}+ \ket{20}\bra{20}) \ea
where $\ket{\psi_\theta} = \cos{\theta}\ket{00}+\sin{\theta}\ket{11}$, where $ b^2 = \sin^2\theta/(2\sin^2\theta+1)$.
Next we test the CHSH inequality on $\rho$, with the following settings: $A_0=\sigma_z$, $A_1=\sigma_x$, $B_0=\sigma_z$, $B_1=\cos{\beta}\sigma_z - \sin{\beta}\sigma_x $ where the Pauli matrices act on the $\{\ket{0},\ket{1}\}$ subspace, and the subspace $\ket{2}$ is mapped to outcome $+1$. With this choice, the separable part of $\rho$ simply gives CHSH=2. The entangled part of $\rho$ gives $\text{CHSH}=1+\sin{\beta}\sin{2\theta}+\cos{\beta}$, which is maximized by setting $\tan{\beta}= \sin{2\theta}$, leading to
\ba \text{CHSH}=1 + \sqrt{1+\sin^2(2\theta)}. \ea
Hence we have that
$ \text{CHSH}(\rho) >2$ for any $0<\theta<\pi/2$. The maximal violation is
\ba \text{CHSH}(\rho)=\frac{1+4\sin^2\theta+\sqrt{1+\sin^2 2\theta}}{1+2\sin^2\theta}. \ea
The maximal value of this expression is $\text{CHSH}=2.2247$, found for $\theta=0.6278$ (corresponding to $b=0.4518$).

It would be interesting to explore the potential of such states in the context of quantum networks \cite{acin}, in which a high persistency of correlations should intuitively be an attractive feature. In particular since the violation of the CHSH inequality implies distillability of entanglement \cite{distill}.

\section{Strength of persistency}

These measures of robustness to losses can be refined. For instance among states of $N$ systems having the same persistency of nonlocality $P_{NL}$, which one is the most robust?
One possibility consists in determining how much white noise must be added to the state in order to decrease its persistency.
More precisely, for a given $N$-party state $\ket{\psi}$ a certain amount $(1-w)$ of maximally mixed state ($\openone/d^N$) is added, so that the resulting state is
\ba \rho= w\ket{\psi}\bra{\psi}+(1-w)\openone/d^N. \ea
We now remove $P_{NL}-1$ systems from $\rho$ from all possible positions. The strength of persistency is the minimum value of $w$ for which all reduced states are still nonlocal.
Here we have estimated the persistency and its strength for $N$-qubit states which are invariant under arbitrary (or cyclic) permutations of the parties. We considered (i) Dicke states, $\ket{D_N^m}$, which are permutationally symmetric (W states correspond to $m=1$), and (ii) translationally invariant states $\ket{T_N^m}$. Formally, one has
\ba \ket{D_N^m} &=& \sqrt{\frac{(N-m)!m!}{N!}} \, \Pi [\ket{0}^{\otimes N-m} \ket{1}^{\otimes m}] \\
\ket{T_N^m} &=& \frac{1}{\sqrt{N}}\Pi_T [\ket{0}^{\otimes N-m} \ket{1}^{\otimes m}]\ea
where $\Pi$ and $\Pi_T$ denote arbitrary and translational permutations.
For each type of states, and up to $N=7$ qubits, we derived lower bounds on $P_{NL}$ by finding an explicit Bell inequality violation (focusing here on Bell inequalities with 2 binary settings per party).
The results are presented in Table~\ref{table}) indicate that $P_E$ and $P_{NL}$ behave quite differently in general, showing that different types of quantum correlations have different robustness to losses. Note that  translationally invariant states $\ket{T_N^k}$ appear to have very robust persistency of nonlocality; also, for $N=6$ qubits, the 2D cluster $\ket{Cl_{2\times3}}$ appears to be the most robust. Moreover, less symmetric states appear to be more robust, suggesting a link between symmetry and persistency, which will be discussed below in Section~VI.

\section{Persistency vs genuine multipartite entanglement}

Although persistency and GME capture different aspects of quantum correlations, it is worth asking how they relate to each other. First note that GME does not imply maximal persistency. Consider for instance the tripartite state, which has genuine tripartite entanglement: $\ket{\phi^+}_{AB}\otimes \ket{\phi^+}_{AC}$, where $\ket{\phi^+}$ denotes a 2-qubit Bell state. Tracing out party $A$ results in a separable state for $BC$.
For pure states, maximal persistency implies GME. This follows from the simple observation that, for any pure non GME state, there are always two parties left in a separable state after tracing out the others. However, for mixed states this link does not hold in general. Consider for instance the tripartite state
\ba \rho &=& \frac{1}{3} \bigg( \ket{0}\bra{0}_A \otimes \ket{\phi^+_{01}}\bra{\phi^+_{01}}_{BC} \nonumber \\ &&
+ \,\, \ket{2}\bra{2}_B \otimes \ket{\phi^+_{23}}\bra{\phi^+_{23}}_{AC} \\\nonumber
&& + \,\, \ket{4}\bra{4}_C \otimes \ket{\phi^+_{45}}\bra{\phi^+_{45}}_{AB} \bigg) \ea
where $\ket{\phi^+_{ij}}= (\ket{ii}+\ket{jj})/\sqrt{2}$.
Although $\rho$ is biseparable (see Eq. \eqref{bisep} below), it has maximal persistency. Each 2-party reduced state violates the CHSH inequality, using the optimal measurements in the relevant qubit subspace, and attributing the outcome +1 in the orthogonal subspace \cite{foot}.

It turns out that persistency and GME can be tested simultaneously. Consider the simple tripartite case, and the following Bell inequality
\ba S = CHSH_{AB}+CHSH_{A'C}+CHSH_{B'C'} \leq 6 \ea
where $A$ and $A'$ denote different sets of measurements for party A, and similarly for parties B and C.
The most general biseparable state is of the form
\ba\label{bisep} \rho_{bs} = pq \rho_{AB|C}  + p(1-q) \rho_{AC|B} + (1-p)(1-q) \rho_{A|BC} \ea
with $0 \leq p,q\leq 1$ and $\rho_{AB|C} = \sum_k{\rho_{AB}^k\otimes \rho_C^k}$. From the fact that $\text{CHSH}_{AB}\leq pq 2\sqrt{2}+(1-pq)2$ (and similar relations for other bipartitions) it follows that
\ba S\leq 4 + 2\sqrt{2} \quad  \text{for all } \rho_{bs}. \ea
If the above inequality is violated, the state is GME. If additionally each CHSH inequality is violated individually, the state has persistency $P_{NL}=2$.
This is the case for instance for a state of the form \eqref{3qutrit} with $a=0.6469$, which gives 
$S=3\times 2.4087> 4 + 2\sqrt 2$.

\section{Persistency vs symmetry}

As mentioned in Section~III, the degree of symmetry of a quantum state seems to affect its persistency.
In particular, translationally symmetric states appear to have stronger persistency compared to permutationally symmetric states. Indeed the 2-party reduced state of a permutationally symmetric $N$-party state $\rho$ cannot violate any Bell inequality with less than $N$ settings, as it admits an $N-1$ symmetric extension \cite{terhal}, given by $\rho$ itself. Hence, although this does not rule out, strictly speaking, the possibility for a symmetric state to have maximal persistency of nonlocality, it nevertheless shows that too much symmetry makes the state less robust against losses, in the sense that revealing the remaining correlations requires a more complicated Bell test. Indeed, more generally, symmetry and independence of subsystems are deeply connected \cite{renato}.

Moreover these ideas allow one to test the asymmetry of an unknown state $\rho$ in a device-independent way, here, from the statistics of measurements on two particles only. In particular, if all 2-party reduced states of $\rho$ turn out to violate a Bell inequality with $N-1$ measurements or less per party, then $\rho$ is not permutationally symmetric. Moreover, this statement can be made quantitative as the smallest Bell violation $S=\text{tr}(\mathcal{B}\rho)>L$ ($\mathcal{B}$ denotes the Bell operator) can then be used to lower bound the trace distance between $\rho$ and its closest symmetric state $\sigma_{\text{sym}}$
\ba D(\rho,\sigma_{\text{sym}})=\frac{1}{2}||\rho-\sigma_{\text{sym}}||_1 \geq \frac{S-L}{2 ||\mathcal{B}||_{\infty}}. \ea
Here we have used the fact that
\ba S-L\leq \text{tr}(\mathcal{B}(\rho-\sigma_{\text{sym}})) \leq || \mathcal{B}||_{\infty} ||\rho-\sigma_{\text{sym}}||_1 \ea
where the last inequality follows from H\"older's inequality; $|| \mathcal{B}||_{\infty}$ is the largest eigenvalue of $\mathcal{B}$.
Note that the trace distance has an operational meaning, and represents the distinguishability of two states under the optimal measurement.
Hence placing device-independent lower bounds on this quantity may be relevant in various contexts.

\section{Conclusion}

We have explored the persistency of multipartite quantum correlations, with a particular focus on nonlocality. This work represents part of a more general research program aiming at characterizing the nonlocal properties of different classes of multipartite entangled states (see e.g. \cite{various,various2}).

The ideas presented here are potentially relevant for several areas of quantum information, in particular for quantum networks.
In this context a high persistency of nonlocality should be an appealing feature, since the violation of certain Bell inequalities implies distillability of entanglement \cite{distill}. Moreover, it would be interesting to study persistency of nonlocality in many-body systems, where persistent entanglement has already proven to be a useful concept \cite{osterloh}. Another potential application worth investigating is a loophole-free multipartite Bell test. Finally, we believe that persistency of nonlocality is relevant experimentally \cite{mohamed}, in particular to provide device-independent tests of persistent entanglement.

\emph{Acknowledgements.} We thank Marcus Huber for suggesting a relevant example in section V. We also acknowledge financial support from the UK EPSRC, the EU DIQIP, the Swiss National Science Foundation (grant PP00P2\_138917), the Hungarian National Research Fund OTKA (PD101461), and a J\'anos Bolyai Grant of the Hungarian Academy of Sciences.

\section{Appendix: States with maximal persistency}


%
Here we show, by constructing an explicit example, that for any odd number of parties $N$, there exists a state $\ket{\Psi_N}$ with local Hilbert space dimension $d=N$, which has maximal persistency of nonlocality, i.e. $P_{NL}(\ket{\Psi_N})=N-1$.

We consider a state of the form
\ba \ket{\Psi_3} &=& a \ket{000} + b \text{ sym} [ \ket{012} ]  \\
 \ket{\Psi_5} &=& a \ket{00000} + b \text{ sym} [ \ket{00012} + \ket{00304} ]   \ea
and so on for any odd $N$. Here $\text{sym}[x]$ means that the expression $x$ must be symmetrized with respect to all cyclic transformations.
Note that the number of terms with coefficient $b$ is $D= \left( \begin{array}{c} N \\ N-2 \end{array} \right)
= N(N-1)/2$.
Each term is such that $N-2$ parties have the state $\ket{0}$. All these terms are orthogonal.
Next we consider the 2-party reduced state of $\ket{\Psi}$, which is given by
\ba \rho = (1- (D-1)b^2) \ket{\psi_\theta}\bra{\psi_\theta}  + (D-1) b^2 \rho_{sep} \ea
where $\rho_{sep}$ is a separable state containing $D-1$ terms which are of 2 possible types:
(i) terms of the form $\ket{0k}\bra{0k}$ (or $\ket{k0}\bra{k0}$) with $k\geq2$; there are $2(N-2)$ of them,
(ii) terms of the form $\ket{00}\bra{00}$; there are $C=D-1 - 2(N-2)$ of them.

We consider the CHSH inequality, with the following choice of settings: $A_0=\sigma_z$, $A_1=\sigma_x$, $B_0=\sigma_z$, $B_1=\cos{\beta}\sigma_z - \sin{\beta}\sigma_x $, where $\tan{\beta}= \sin{2\theta}$. With this choice of settings, one obtains the following CHSH values for the different components of the state $\rho$:
\ba \text{CHSH}(\ket{\psi_\theta}) &=&   1 + \sqrt{1+\sin^2(\theta)} \\
\text{CHSH}( \ket{0k}\bra{0k} ) &=&   2 \\
\text{CHSH}(\ket{00}\bra{00}) &=&   1 + \frac{1}{\sqrt{1+\sin^2(\theta)}}
\ea
The CHSH value of $\rho$ is then
\ba \text{CHSH}(\rho) &=&  (1- (D-1)b^2) \text{CHSH}(\psi_\theta) \\\nonumber
& &+ Cb^2 \text{CHSH}(\ket{00}\bra{00})    + 4(N-2)b^2 \ea
For convenience, let us redefine the value of CHSH such that the local bound is 0 (i.e. simply subtracting the constant 2).
In order to have Bell violation, we require
\ba \text{CHSH}(\rho) &=& (1- (D-1)b^2) \text{CHSH}(\psi_\theta) \\ \nonumber & &+ Cb^2 \text{CHSH}(\ket{00}\bra{00}) > 0 \ea
which gives
\ba (1- (D-1)b^2) X + Cb^2 /X  >  (1- (D-1)b^2) +Cb^2 \nonumber \ea
where $X=\sqrt{1+\sin^2(2\theta)}$.
Here we will focus on the regime where $\theta$ is small, hence we can do the approximations $X = 1+2\theta^2$ and $1/X = 1-2\theta^2$.
With this, the condition for CHSH violation becomes
\ba    \theta^2 (1 - (D-1+C)b^2) >0    \ea
which holds for $b$ sufficiently small.

Finally, note that, although the above construction works for odd $N$, we believe that states with similar properties can be found for even $N$ as well. For the case $N=4$, we found the following translationally invariant state
$\ket{\psi_4}=0.3039(\ket{0112}+\ket{1120}+\ket{1201}+\ket{2011})+0.2566(\ket{0202}+\ket{2020})
-0.3033(\ket{1313}+\ket{3131})+0.4783\ket{1111}+0.2563\ket{3333}$. This state violates
the CHSH inequality by an amount of $CHSH=2.3226$ after removing 2 qubits from sites (A,B), (B,C), (C,D), (D,A), whereas by removing (A,C) or (B,D), it violates CHSH up to $\text{CHSH}=2.3216$.


\begin{thebibliography}{99}
\bibitem{foot} This example was kindly suggested to us by Marcus Huber.
\end{thebibliography}
\end{document}